\documentstyle [aps,prd]{revtex}
\input epsf
\begin{document}

\newcommand{\xe}{$x_e~$}
\newcommand{\tdot}{$\dot{\tau}~$}
\newcommand{\be}{\begin{equation}}
\newcommand{\ee}{\end{equation}}
\newcommand{\obh}{$\Omega_b h^2~$}
\newcommand{\omh}{$\Omega_m h^2~$}
\newcommand{\och}{$\Omega_c h^2~$}
\newcommand{\okh}{$\Omega_K h^2~$}
\newcommand{\olh}{$\Omega_\Lambda h^2~$}
\newcommand{\omrdh}{$\Omega_{rd} h^2~$}
\newcommand{\omb}{$\Omega_b~$}
\newcommand{\omm}{$\Omega_m~$}
\newcommand{\omc}{$\Omega_c~$}
\newcommand{\omk}{$\Omega_K~$}
\newcommand{\oml}{$\Omega_\Lambda~$}
\newcommand{\omrd}{$\Omega_{rd}~$}
\newcommand{\et}{$\tau~$}
\newcommand{\ph}{$\phi$}
\newcommand{\ps}{$\psi$}
\newcommand{\phd}{\dot{\phi}}
\newcommand{\psd}{\dot{\psi}}
\newcommand{\potd}{$\dot{\phi}+\dot{\psi}~$}
\newcommand{\ddt}{\frac{\partial}{\partial \tau}}
\newcommand{\hdot}{\dot{h}}
\newcommand{\etadot}{\dot{\eta}}
\newcommand{\epsn}{\epsilon}
\newcommand{\mpc}{\mathrm{Mpc}}

\newcommand{\deltard}{\delta_{rd}}
\newcommand{\thetard}{\theta_{rd}}
\newcommand{\shearrd}{\sigma_{rd}}
\newcommand{\rhoprd}{\rho_{rd}+P_{rd}}
\newcommand{\rrddot}{\dot{r}_{rd}}
\newcommand{\adotoa}{\frac{\dot{a}}{a}}
\newcommand{\adotoasq}{\left(\frac{\dot a}{a}\right)^2}
\newcommand{\order}[1]{\ensuremath{\mathcal{O}(#1)}}

\newenvironment{comm}{\begin{list}{}{\itshape \item}}
{\end{list}}

\title{Improved Treatment of Cosmic Microwave Background Fluctuations Induced by a Late-decaying
Massive Neutrino}
\author{Manoj Kaplinghat,$^{1}$ Robert E. Lopez,$^{2}$
Scott Dodelson,$^3$ and Robert J. Scherrer$^{1,4}$}
\address{$^1$Department of Physics, The Ohio State University,
Columbus, OH~~43210}
\address{$^2$Department of Physics,
Enrico Fermi Institute, University of Chicago, Chicago, IL~~60637-1433}
\address{$^3$NASA/Fermilab Astrophysics Center,
Fermi National Accelerator Laboratory, Batavia, IL 60510}
\address{$^4$Department of Astronomy, The Ohio State University,
Columbus, OH~~43210}
\date{\today}
\maketitle
\begin{abstract}
A massive neutrino which decays after recombination ($t \ge 10^{13}$ sec) into
relativistic decay products produces an enhanced integrated Sachs-Wolfe effect,
allowing constraints to be placed on such neutrinos from present cosmic microwave
background anisotropy data.  Previous treatments of this problem have
approximated the decay products as an additional component of
the neutrino background.  This approach violates energy-momentum conservation,
and we show that it leads to serious errors for some neutrino
masses and lifetimes.  We redo this calculation more accurately,
by correctly incorporating the spatial distribution of the decay products.
For low neutrino masses and long lifetimes, we obtain a much
smaller distortion in the CMB fluctuation spectrum than have previous
treatments.  We combine these new results with a recent set of
CMB data to exclude the mass and lifetime range $m_h > 100$ eV,
$\tau > 10^{12}$ sec.  Masses as low as 30 eV are excluded for a narrower
range in lifetime.
\end{abstract}

\section{Introduction}
Anisotropies in the cosmic microwave background
(CMB) contain an enormous amount of information about the universe. 
Data presently available
has been used\cite{Bunn}-\cite{MAX} to constrain from two up to eight cosmological parameters.
With the promise of ever more precise measurements of these anisotropies, 
it has become possible to envision CMB fluctuations as a tool to go beyond this
minimal set and
constrain other areas of physics.  Recent proposed constraints include limits on
Brans-Dicke theories \cite{Liddle}, constraints on time-variation in the
fine-structure constant \cite{Hannestad,alpha}, tests of finite-temperature
QED~\cite{QED}, and limits on various models for both stable\cite{stable} and
unstable \cite{unstable,lopez,hannestad} massive neutrinos.  All of these additional
constraints, with one exception, are based on the high-precision fluctuation spectra
expected from the MAP and PLANCK satellites. The sole exception is
reference~\cite{lopez}, in which Lopez et al. pointed out that the radiation from a
neutrino decaying into relativistic decay products could produce such a large
integrated Sachs-Wolfe (ISW) effect, that a fairly large mass-lifetime range can be
ruled out from current observations.  Lopez et al. argued that a neutrino with a mass
greater than 10 eV and a lifetime between $10^{13}$ and $10^{17}$ sec could be ruled
out.  (Although this calculation assumes nothing about the nature of the decay
products other than that they are relativistic, this limit is most useful when
applied to decay modes into ``sterile" particles such as a light neutrino and a
Majoron, since other, more restrictive limits apply to photon-producing decays).
Hannestad~\cite{hannestad} showed that the MAP and PLANCK experiments should produce
an even larger excluded region in the neutrino mass-lifetime plane.

In this paper, we improve on a major approximation of
references~\cite{lopez,hannestad}. In these papers, the relativistic decay products
were simply added to the background neutrino energy density in the program CMBFAST
\cite{selj+zald}.  However, when the massive neutrinos decay, the spatial
distribution of the decay products is determined by the distribution of the
non-relativistic decaying particles; it is not identical to the distribution of the
background massless neutrinos. In fact, the approach of references
\cite{lopez,hannestad} violates energy-momentum conservation.  Although in this
approach energy and momentum are explicity conserved at zeroth order (the mean), the
first order perturbations violate energy-momentum conservation. This may seem like a
small effect, but it actually has significant consequences for the CMB fluctuation
spectrum.

In the next section, we discuss the formalism for the Sachs-Wolfe effect in the
presence of neutrinos decaying after recombination.  In section III, we present our
results, showing the effects of correctly incorporating the spatial distribution of
the decay products, and provide a simple physical explanation of these effects.  In
section IV, we show how our revised calculation affects the excluded region in the
neutrino mass-lifetime plane, and in section V we briefly summarize our
conclusions.  A comparison of
our new results with current data leads to the excluded region $m_h > 100$ eV,
$\tau > 10^{12}$ sec, although smaller masses can also be excluded for a smaller
range of $\tau$.

\section{The ISW Effect with an Unstable Neutrino:  Formalism}
To calculate the CMB fluctuations in the presence of a decaying massive neutrino, we
first review the basic precepts of the pertinent linear perturbation theory. The
perturbed homogeneous, isotropic FRW metric can be parametrized as
\be ds^2=-a(\tau)^2\left[d\tau^2(1+2\psi)-d\vec{x}^2(1-2\phi)\right], 
\label{metric} \ee
where $a$ is the scale factor normalized to unity today and \et is the conformal time
defined by $d\tau=dt/a$, $t$ being the proper time of a comoving observer. This
particular gauge is referred to as the conformal Newtonian gauge because the behavior
of the potentials (\ph, \ps) is akin, loosely speaking, to that of the Newtonian potential. These
potentials determine the large scale CMB behavior. In particular, the photon
temperature perturbation decomposed into its Fourier and angular modes can be shown
to be~\cite{hu+sugy}
\be \Delta_{\ell}(k)=\int_0^{\tau_0} d\tau \, 
(\phd(k, \tau)+\psd(k, \tau))\exp(-\kappa(\tau)) j_{\ell}(k\tau_0-k\tau),
\label{delk} \ee where the subscript `0' refers throughout to the present time and
$\kappa$ is the optical depth from the present to some conformal time \et in the
past. For the purpose of clarity, all the sources contributing to the anisotropy from
inside the last scattering surface have been set to zero in Eq.~\ref{delk}. The
effect of the sources contributing to the anisotropy between last scattering and the
present (as given in Eq.~\ref{delk}) is called the Integrated Sachs-Wolfe (ISW)
effect. The power in the $\ell^{th}$ multipole is normally defined as $\ell
(\ell+1)C_\ell$ with~\cite{selj+zald}
\be C_\ell = (4\pi)^2 \int_0^\infty dk\, k^2 |\Delta_{\ell}(k)|^2.
\label{cl} \ee

The sources mentioned in connection with the ISW effect can be varied. At any time,
the modes that are important to the ISW effect correspond to those scales which are
smaller than the sound horizon of the whole fluid (matter+radiation) at that
time. For these modes, the potentials can decay if there is radiation pressure or if
the universe expands rapidly. In models with no cosmological constant, the main
contribution to the ISW effect comes from just after recombination (since radiation
redshifts faster than matter). Inclusion of a cosmological constant leading to a
rapid expansion of the universe late in its history would boost the power on larger
scales (small $\ell$). Any other astrophysical process which contributes to the
radiation content of the universe between last scattering and the present will lead
to an increase in the total ISW effect. One such scenario is that of a massive
particle decaying around or after last scattering. We will consider the case of a
massive neutrino decaying non-relativistically into (effectively) massless
particles. The details of the daughter particles turn out to be irrelevant.

To quantify the evolution of the massive neutrino density, we will consider the
Boltzmann equation for its distribution. In a homogeneous and isotropic universe, the
distribution of the collisionless massive neutrino decaying non-relativistically into
two massless particles follows~\cite{kawasaki}
\be \ddt f^0_h (q_h, \tau) = -\frac{a^2 m_h}{t_d \epsn_h} f^0_h (q_h, \tau), 
\label{dfhdt0} \ee
where $t_d$ is the mean lifetime of the neutrino, and $\epsn_h$
and $q_h$ are the comoving energy and momentum:
\(\epsn_h^2 =q_h^2+m_h^2a^2\), and a superscript `0' will be used
throughout to denote unperturbed quantities.
We make the following simplifications throughout our treatment: (1)
neglect inverse decays, (2) neglect spontaneous emission, (3) neglect Pauli blocking
factor. The solution approaches the familiar $\exp(-t/t_d)$ behavior as the neutrino
becomes non-relativistic. The evolution equation for the energy density of the
unstable neutrino is the integral of Eq. \ref{dfhdt0}. It reads
\be \dot{\rho}^0_h+3\frac{\dot{a}}{a}(\rho^0_h+P^0_h)=
-\frac{am_hn^0_h}{t_d}, \label{rhohdot} \ee where overdots represent differentiation
with respect to conformal time. It should be noted (as it is important if the decay
is not completely non-relativistic) that the right hand side contains the product of
$m_h$ and $n^0_h$ (number density) and not $\rho^0_h$. 

We now turn on the perturbations in the metric. Although the conformal Newtonian
gauge is the most useful in which to understand the ISW effect, for
computational purposes\footnote{The main advantage is that CMBFAST 
\cite{selj+zald} is written in synchronous gauge.} we will define all our
variables in the synchronous gauge.  Thus, we will express the integrand
in Eq. 2 in terms of perturbations in the synchronous gauge.
The synchronous gauge
has the property that the coordinate time and the proper time of a freely falling
observer coincide. All the perturbations are in the spatial part of the metric
($g_{ij}=a^2\delta_{ij}+a^2h_{ij}$) in this gauge. The perturbation $h_{ij}$ can be
Fourier transformed and broken up into its trace and a traceless part
as~\cite{ma+bert}
\be 
h_{ij}(\vec{x}, \tau)=\int d^3k\, 
\frac{\exp(i\vec{k}\cdot\vec{x})}{k^2}
\left[h(\vec{k},\tau)k_ik_j+6\eta(\vec{k},\tau)
\left(k_ik_j-\frac{k^2}{3}\delta_{ij}\right)\right]. \label{defh} \ee
Instead of working with the conjugate momentum in the perturbed space-time, we will
use $q_h$ and $\epsn_h$ as defined above~\cite{bond+szal} and in keeping with that,
we will write out the perturbed massive neutrino distribution as
\be 
f_h (\vec{x}, \vec{q}_h, \tau) = f^0_h(q_h, \tau)
\left[ 1 + \Psi_h (\vec{x}, \vec{q}_h, \tau)\right]
 \,. \label{fh} 
\ee 
Due to the fact that the decay term is linear in $f_h$, the form of the equation for
the evolution of $\Psi_h$ is identical to that of the stable massive neutrino but
with $f^0_h$ now given by Eq.~\ref{dfhdt0}.  The stable massive neutrino case has
been clearly worked out in Ref.~\cite{ma+bert}.

The decay radiation rises exponentially from being negligible in the past 
to some maximum value at $\tau\sim \tau_d$ and then drops off as $a^{-4}$ like 
normal radiation. It is more informative therefore to follow the quantity 
\(r_{rd}=\rho^0_{rd}/\rho^0_\nu\) where `rd' denotes the decay radiation and
$\rho_\nu^0$ is the cosmological density in a massless neutrino. The evolution equation
for $r_{rd}$ is
\be \dot{r}_{rd}=\frac{m_hn^0_h}{\rho^0_\nu}\frac{a}{t_d}. \label{rrddot}\ee 
The treatment of the perturbations in the decay radiation will be analogous to that
of the massless neutrino as worked out in Ref.~\cite{ma+bert}.  To evolve the
perturbations in the decay radiation, we will integrate out the momentum dependence
in the distribution function by defining (in Fourier space)
\be F_{rd}(\vec{k}, \hat{n}, \tau)
=\frac{\int dq\, q^3f^0_{rd}(q, \tau)\Psi_{rd}(\vec{k}, q, \hat{n}, \tau)}
{\int dq\, q^3 f^0_{rd}(q, \tau)}\;r_{rd}, \label{frd} 
\label {eq:be}
\ee
where $\vec{q}=q\hat{n}$ and $\Psi_{rd}$ is defined analagously to
Eq. \ref{fh}. The equation governing the evolution of $F_{rd}$ can be
worked out to give
\begin{eqnarray} 
\dot{F}_{rd}+ i k\mu F_{rd}+4\left(\frac{\dot{h}}{6}+
\frac{\hdot+6\etadot}{3}
P_2(\mu)\right)r_{rd}&=&\dot{r}_{rd}\left(N_0-3 i
N_1P_1(\mu)-\frac{2}{3}N_2P_2(\mu)+\ldots\right), \label{frddot}\\
N_0(k, \tau)&=&     
\frac{\int dq_h\, q_h^2f^0_{h}(q_h, \tau)\Psi_{h}(k, q_h, \tau)
\left(1-\frac{8}{3}\left(\frac{q_h}{am_h}\right)^2+\ldots\right)}
{\int dq_h\, q_h^2 f^0_{h}(q_h, \tau)}, \label{N_0} 
\end{eqnarray}
% \begin{comm}... added the ellipsis to eq.~\ref{frddot}.
% Reshuffled the next para to make it more coherent.
% And a typo (${\cal O}(...)$ part) which preplexed Rob+Scott has been
% un-typo-ed.\end{comm}
where $\mu=\hat{k}\cdot\hat{n}$ and $P_n(\mu)$ are the Legendre polynomials of order
$n$.  The series of terms in these equations arises because the perturbed quantities depend on the
direction of momentum and to get the contribution to a daughter particle with
momentum $\vec{q}$, we need to integrate over all possible $\vec{q}_h$.  Thus
Eq.~\ref{frddot} depends on both $\mu$ and $\vec{q}\cdot\vec{q}_h$. The situation
simplifies enormously for non-relativistic decays because each term $N_p$, which
contributes to the $p^{th}$ multipole progressively, is of ${\cal O}(\langle q^p_h
\rangle /a^pm^p_h)$ or higher.  In Eqs.~\ref{frddot} and~\ref{N_0}, the series has
been truncated by only keeping terms up to ${\cal O}(q^2_h/a^2m^2_h)$ in the
integrand.  Similar equations for the evolution of perturbations in the decay
radiation can be found in references~\cite{bond+efst},~\cite{bhar+seth}.
Apart from $N_0$, the terms on the right-hand side of Eq.~(\ref{frddot}) are
completely negligible for non-relativistic decays. 

The use of Eq.~\ref{frddot} is our only difference from the treatment in
ref. \cite{lopez}.  In the latter paper, the relativistic decay products were simply
added to the neutrino background in CMBFAST.  This is equivalent to setting the right
hand side of Eq. 10 to zero. Since the perturbations in the decay products are
determined by the perturbations in both the metric and the decaying massive
particles, they are correctly described by Eq.~\ref{frddot}.  Although this may seem
like a minor difference, it produces very large effects, as we now show.

\section{The ISW Effect with an Unstable Neutrino:  Results}

The formalism outlined above for the evolution of an unstable neutrino and its decay
products was integrated into the CMBFAST code~\cite{selj+zald}. We investigated a
range of masses from 10 eV to $10^4$ eV and lifetimes from $10^{12}$ to $10^{18}$
seconds. The underlying cosmology was taken to be a standard ($\Omega=1$) CDM model
with $h = 0.5$ (with $H_0 = 100h$ km sec$^{-1}$
Mpc$^{-1}$); baryon density $\Omega_B h^2 = .02$
and scale invariant isentropic initial conditions
(the same model was used in ref.
\cite{lopez}).
Our results are shown in Fig. 1 for several masses and lifetimes, along with the
results obtained by simply adding the decay products to the relativistic
background. As pointed out in Ref. \cite{lopez} there is indeed an enhancement
in the spectra at relatively large scales due to the ISW effect produced by the
decaying neutrino. We will see in section IV that for many values of neutrino
mass and lifetime, the spectrum produced is far from that observed today, and therefore
a large region of parameter space is ruled out due to this effect.

\begin{figure}[Fig1]
\centering
\leavevmode\epsfxsize=12cm \epsfbox{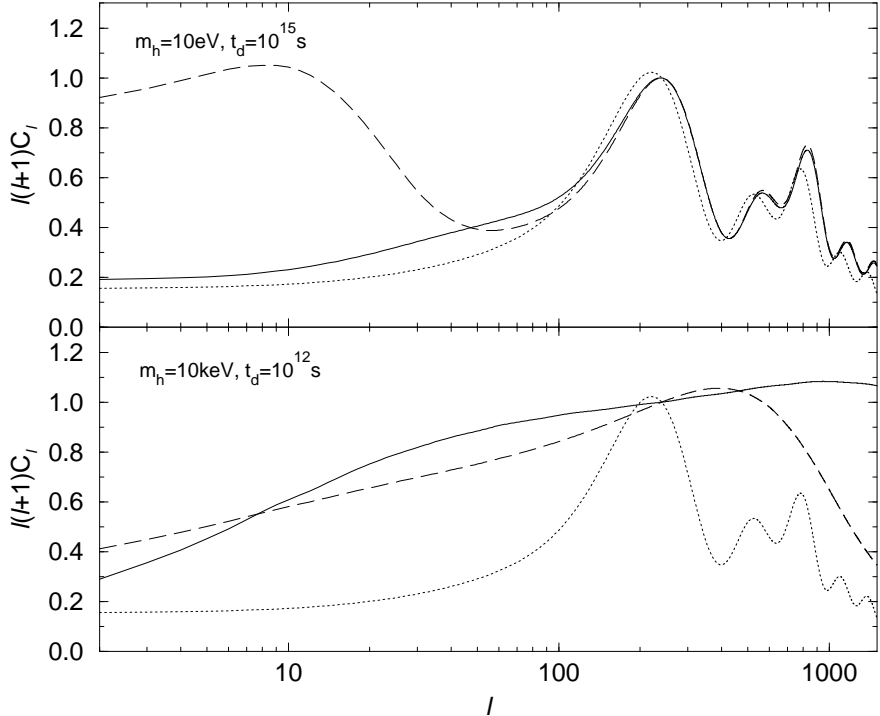}\\
\
\caption[Fig1]{\label{clfig}  The spectrum of CMB fluctuations for
a standard CDM model ($\Omega_b = 0.08$, $h = 0.5$) in the presence of a decaying
neutrino with the indicated mass and lifetime (solid curve).  Shown for comparison is
the spectrum obtained in Ref.~\cite{lopez} where the decay products were added to the
background neutrinos (dashed curve).  The dotted curve gives the fluctuation spectrum
in the absence of a decaying neutrino.}
\end{figure} 

The location of this ISW induced bump is determined by the lifetime of
the neutrino. For lifetimes shorter than the age of the universe, inhomogeneities
on scales $k$ project onto angular scales $\ell \sim k\tau_0$ where $\tau_0$
is the conformal time today, and we assume a flat universe. The potentials
vary in time (and hence cause the ISW effect) most significantly at the
time of decays on scales of order the sound horizon: 
$k_{sh}^2 \simeq 3/(4\tau_d^2 w)$
where $w = P/\rho$. Therefore, the bump in the spectrum is produced
at $\ell \sim k_{sh} \tau_0 \simeq (\tau_0/\tau_d)(4w/3)^{-1/2}$.
At these late times, the dominant contribution to $w$ comes from the
decay radiation; hence $w \simeq\Omega_{rd}/3$ where $\Omega_{rd}$ is
the fraction of critical density in decay radiation. Therefore, the
ISW bump should be roughly at
\be
\ell_{ISW} \simeq \frac{\tau_0}{\tau_d} \,
\sqrt{\frac{9}{4\Omega_{rd}(\tau_d)}} \,.
\ee
For a matter dominated universe the conformal time and time are related as follows:
$\tau \propto t^{1/3}$. For a $m_h = 10$ eV, $t_d = 10^{15}$ sec neutrino,
$\Omega_{rd} \simeq 0.15$ and $\tau_0 / \tau_d \simeq (4\times 10^{17} \,\rm{sec}/
10^{15}\,\rm{sec})^{1/3} \simeq 7.4$. Therefore, in this case we expect 
$\ell_{ISW}\simeq 29$. The actual peak occurs at a
larger value of $l$, due to entropy fluctuations which decrease $w$, 
thereby increasing $k_{sh}$ and, finally, $\ell_{ISW}$.
%\begin{comm}... adiabatic (long modes) $c_s^2=\dot{P}/\dot{\rho}=4/3\omega$.
%ISW peak at 55: though it looks like (really??) its 50 from the plot, if you 
%plot just the ISW power spectrum, it peaks at about 55.\end{comm}

Notice from Figure 1 that we find quantitative disagreement with the results of Lopez
et al. \cite{lopez} (dashed curves).  The new results show that a more accurate
treatment of the spatial distribution of the decay products produces a surprisingly
large change in the CMB fluctuation spectrum compared to the
results of reference \cite{lopez}.  This difference is larger for smaller
masses as can be seen in the figure.

% Height of ISW peak is diminished when approximation is relaxed...
At least for low masses, the most obvious difference between
the the old and new spectra is the smaller size of the ISW effect for the new
case. This difference has a physical explanation: by not properly treating the
perturbations in the decay radiation we overestimate an important source of the
potential decays that drive the ISW effect. To see this we first expand the
Boltzmann equation for decay radiation perturbations, Eq.~\ref{eq:be}, in multipole
moments, $F_{rd} = \sum_l F_{rd,l} \,P_l$, to obtain the following hierarchy, shown
here for $l\le 2$:
\begin{eqnarray}
\dot{\delta}_{rd}+\frac{2}{3}\left(\dot{h}+2\thetard\right)&=&
\frac{\rrddot}{r_{rd}}\left(\delta_h-\deltard\right) , \label{deltarddot}\\
\dot{\theta}_{rd}-k^2\left(\frac{\deltard}{4}-\shearrd\right)&=&
-\frac{\rrddot}{r_{rd}} \thetard , \label{thetarddot}\\
\dot{\sigma}_{rd}-\frac{2}{15}\left(2\thetard+\hdot+6\etadot\right)&=&
-\frac{\rrddot}{r_{rd}} \shearrd , \label{shearrddot}
\end{eqnarray}
where $\deltard=F_{rd,0}/r_{rd}$, $\thetard=3kF_{rd,1}/4r_{rd}$ and
$\shearrd=F_{rd,2}/2r_{rd}$. The treatment of Lopez et al.~\cite{lopez} is equivalent
to neglecting the right hand sides of the equations above. This simplification breaks
down near $\tau \sim
\tau_d$, where $\rrddot/r_{rd}$ is not negligible. 

\begin{figure}[hb]
\centering
\leavevmode\epsfxsize=12cm \epsfbox{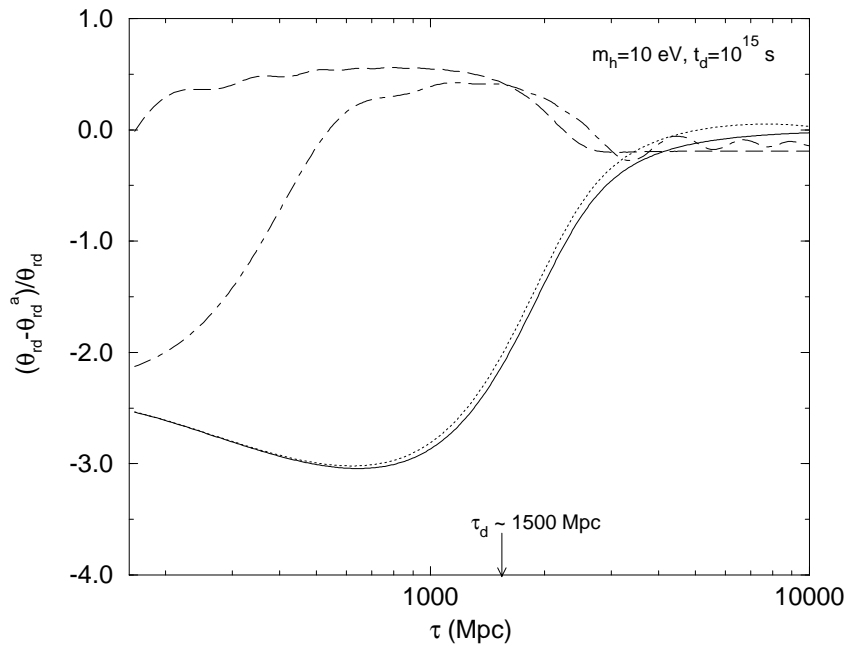}\\
\
\caption[]{\label{dtheta}%  
The difference in the variable $\theta_{rd}$ (dipole of the
decay-produced radiation) between the correct and
the approximate treatment in Ref.~\cite{lopez} is plotted for different
modes: $k^{-1}=2.8\times 10^4$ Mpc (solid curve), $1.4\times 10^3$ Mpc
(dotted curve), 73 Mpc (dot-dashed curve) and 23 Mpc (dashed curve).}
\end{figure}

Neglecting the $\dot r_{rd} / r_{rd}$ terms in the Boltzmann equations for the decay
radiation perturbations results in errors in the perturbations.  Let us focus on
$\theta_{rd}$, which turns out to be primarily responsible for the big difference.
Consider Eq. \ref{thetarddot} for modes above the horizon at $\tau\sim \tau_d$, since
for these modes the approximate treatment of Ref.~\cite{lopez} gives wrong
results. For these modes the $k^2$ terms calculated in the approximate scheme can be
shown (see Appendix) to be roughly similar to its exact value.  Then 
the exact solution $\theta_{rd}$ is related to the
approximate solution $\theta_{rd}^a$ by
\begin{equation}
\dot{\theta}_{rd} \simeq \dot{\theta}^a_{rd} 
- \frac{\dot r_{rd}}{r_{rd}}\,\theta_{rd}
\,.
\end{equation}
%\begin{comm}... the $k^2$ terms cannot be ignored. 
%$\delta, \sigma\sim k^2\tau^2$  while $\theta \sim k^4\tau^3$. \end{comm}
where the superscript here and in what follows denotes the solution to the set
of equations \ref{deltarddot}-\ref{shearrddot} obtained in the approximate 
scheme by neglecting the feedback terms on the right hand side. 
The exact solution for $\thetard$ is therefore much smaller than the approximate one.
Examples for several different modes are shown in 
Figure~\ref{dtheta}. 

These large overestimates of $\theta_{rd}$ lead to correspondingly large
overestimates of the ISW effect and are primarily responsible for the differences
between our spectra and those generated in Ref. \cite{lopez}. The Appendix
demonstrates precisely how the perturbations in the decay-produced radiation affect
the potentials that govern the ISW effect, and how treating the decay products as
identical to the massless neutrinos violates energy-momentum conservation.  The
bottom line is that the ISW effect depends significantly on the behavior of
$\theta_{rd}$ and inaccuracies in it lead directly to inaccuracies in the $C_l$'s.
%\begin{comm} ... all 3 terms contribute the same order of magnitude.
%Infact 2 of those are opposite in sign to the $\theta_{rd}$ term and
%hence cancel significantly to give the eventual ISW effect.
%So it is perhaps a bit strong to say "primarily driven". \end{comm}  
Why does the approximation work better for higher mass neutrinos?
The ISW effect is generated during times when the universe has
appreciable radiation. For low-mass neutrinos whose decay radiation
never dominates the energy density, the decay radiation redshifts away
relative to the matter, and is only important near $\tau\sim \tau_d$.
Therefore, neglecting the $\rrddot/r_{rd}$ terms creates errors in the
decay radiation perturbations at the crucial time when they are driving
the ISW effect.
% The height reduction is less for larger masses...
If the neutrino is massive enough, then its decay products are important for a range
of times with $\tau \gg \tau_d$ when the approximation is good. So the
approximate treatment works better for higher-mass neutrinos, like the $m_h = 10$
keV, $t_d = 10^{12}$ sec case.

There are other visible differences between the anisotropy spectra generated
in reference \cite{lopez} and our more accurate treatment.
One difference, which exacerbates the rise in power at large
scales, is a drop in the small-scale ISW effect. For modes which enter the horizon
when there is significant radiation, the $\delta_h$ term in Eq.~(\ref{deltarddot}) is
an important source term. This increases $\deltard$ relative to $\deltard^a$ and
since $\deltard$ is a source for the evolution of $\thetard$, it implies that
$\thetard^a < \thetard$.  Thus there is a decrease in the ISW effect at small scales
in the approximate scheme of ref \cite{lopez}. This is not visible for the 10 eV
unstable neutrino (in Fig.~\ref{clfig}) because of the comparatively large signature
of the first peak, but it is readily apparent for the 10 keV neutrino because of the
large ISW effect at small scales.

\section{Comparison with current CMB data}

Since the detection of anisotropies in the CMB by COBE\cite{cobe},
there have been dozens of observations of anisotropies on a wide
variety of angular scales (refs. \cite{firs}-\cite{cat}). We now use these
observations to place more accurate limits on neutrino mass and lifetime. 

In ref. \cite{lopez}, a very rough constraint was placed on
decaying neutrino models:  a model was excluded if
the power at $l=200$ was greater than at $l=10$.  As we have
noted in the previous section, a more accurate treatment
of the decaying neutrinos results in a much
smaller distortion in the CMB spectrum for a certain
range of neutrino masses and lifetimes.  However, as we will
see, consideration of all the data leads to constraints which are
almost as stringent as the rough contours in ref. \cite{lopez}. 

CMB experiments typically report an estimate of the band power 
\be
\hat C_i = {1\over 4\pi} { \sum_l (2l+1) W_{i,l} C_l \over 
	\sum_l W_{i,l}/l}
\ee
where $W_{i,l}$ is the window function which depends on beam size and
chopping strategy of experiment $i$. 
Each of these comes with an error bar or, in the case of
correlated measurements, an error matrix $M^{-1}$. The naive way to constrain 
parameters in a theory then is to form
\be
\chi^2 = \sum_{i,i'} \left( \hat C_i - C_i(C_l) \right) M_{ii'}
		\left( \hat C_{i'} - C_{i'}(C_l) \right)
.\ee
Here we have explicitly written the dependence of $C_i$ on the
theoretical $C_l$'s which in turn depend on the cosmological
parameters. This naive statistic is useful only if the band power errors
are Gaussian. In fact, the probability distribution is
typically non-Gaussian, with a large tail at the high end and
a sharp rise at the low end of the distribution. In recognition 
of this, and guided by some compelling theoretical arguments,
Bond, Jaffe, and Knox \cite{BJK} proposed forming
an alternative statistic:
\be
\chi^2 = \sum_{i,i'} \left( \hat Z_i - Z_i(C_l) \right) M^Z_{ii'}
		\left( \hat Z_{i'} - Z_{i'}(C_l) \right)
\label{chisq}\ee	
where 
\be
Z_i \equiv \ln\left( C_i + x_i\right)
\ee
with $x_i$ an experiment dependent quantity, determined by the noise.
The covariance matrix is now
\be
M^Z_{ij} = \left( \hat C_i + x_i \right)  M_{ij} \left( \hat C_i + x_i \right)
.\ee
Bond, Jaffe, and Knox \cite{BJK} have tabulated and made available
the relevant data from the experiments in refs. \cite{cobe}-\cite{cat}.
We use this information and formalism\footnote{We also account for
calibration uncertainty in the manner set down in ref. \cite{BJK}.}
to constrain the mass and lifetime
of unstable neutrinos.

\begin{figure}[Fig3]
\centering
\leavevmode\epsfxsize=12cm \epsfbox{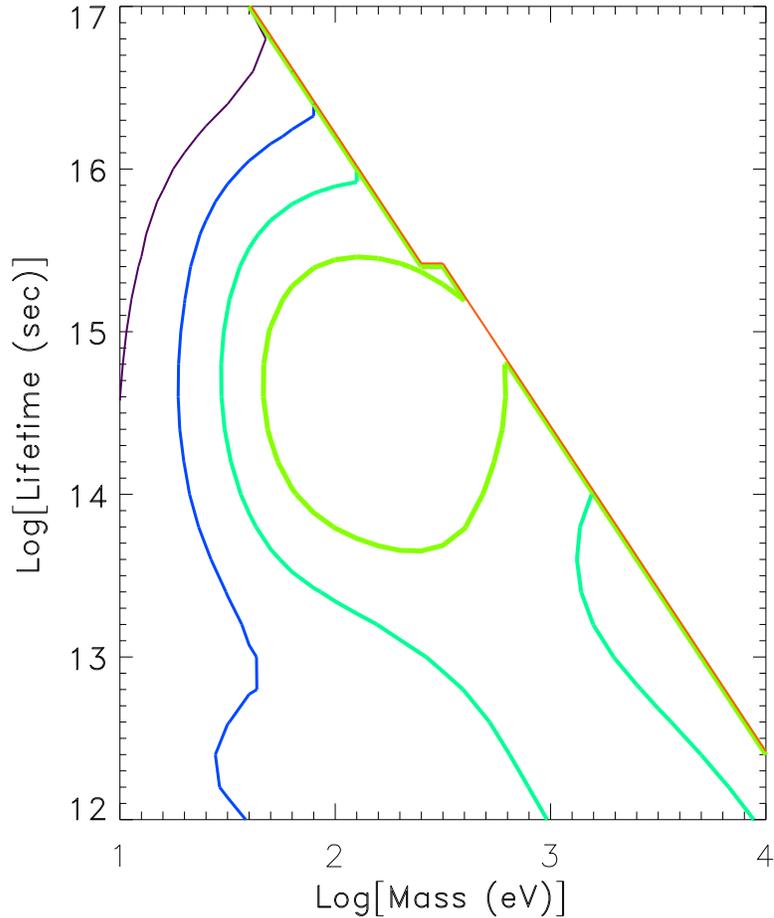}\\
\
\caption[Fig3]{\label{cont}  Contours of $\chi^2$ in neutrino mass/lifetime
plane in $1-\sigma$ intervals. Inner contour is ruled out at $4\sigma$; outermost contour is
$1\sigma$.
The upper right part of the figure leads to
$\Omega > 1$ (the jaggedness at $\log\tau \simeq 15.5$ reflects the grid size
used to explore the parameter space). Here $h=0.5;\Omega_B=0.08$.}
\end{figure}

The $\chi^2$ in eq. \ref{chisq} depends on the parameters of the
cosmological model. In principle, it would be nice to allow
as many parameters as possible to vary in addition to the mass
and lifetime of the neutrino. This must be balanced against the
constraints imposed by non-negligible time needed to run the
modified version of CMBFAST\footnote{The modified version, accounting
for decaying neutrinos, takes about ten times longer than the
plain vanilla code.}. Our strategy is to vary the mass and lifetime
of the neutrino; the overall normalization of the $C_l$'s; the
primordial spectral index (equal to one for Harrison-Zel'dovich 
fluctuations); and the calibration of each experiment. For the other
cosmological parameters, we make ``conservative'' choices. That is, we choose
values likely to make the power on small scales $(l\sim 200)$ as large
as possible compared with the power on large scales. This acts against
the effect of the decaying neutrino, which boosts up power on large scales,
and therefore leads to more conservative limits. At each point in 
$(m,\tau)$ space, we use a Levenberg-Marquardt algorithm (see e.g. \cite{Press,DK})
to find the
values of normalization, spectral index, and calibration which
minimize the $\chi^2$ defined in Eq. \ref{chisq}. The contours in Fig. \ref{cont}
show these best fit $\chi^2$ in the $(m,\tau)$ plane.

Figure \ref{cont} shows the constraints on the neutrino
mass and lifetime for a Hubble constant $h = 0.5$ and
$\Omega_B h^2= .02$ in a flat ($\Omega=1$)
matter dominated ($\Omega_\Lambda=0$) universe. 
The high baryon content
is above the favored value of Tytler and Burles \cite{TB} and serves to
raise the power on small scales. Masses greater than $100$ eV are ruled out 
for almost all lifetimes we have explored ($\tau > 10^{12}$ sec). For lifetimes
between $10^{14}$ and $10^{15}$ sec, masses as low as $30$ eV are excluded at
the two-sigma level. These results are similar to those of ref. \cite{lopez},
but more reliable because of the improvements in the calculated spectra and
the more careful treatment of the data. 

\begin{figure}[Fig4]
\centering
\leavevmode\epsfxsize=12cm \epsfbox{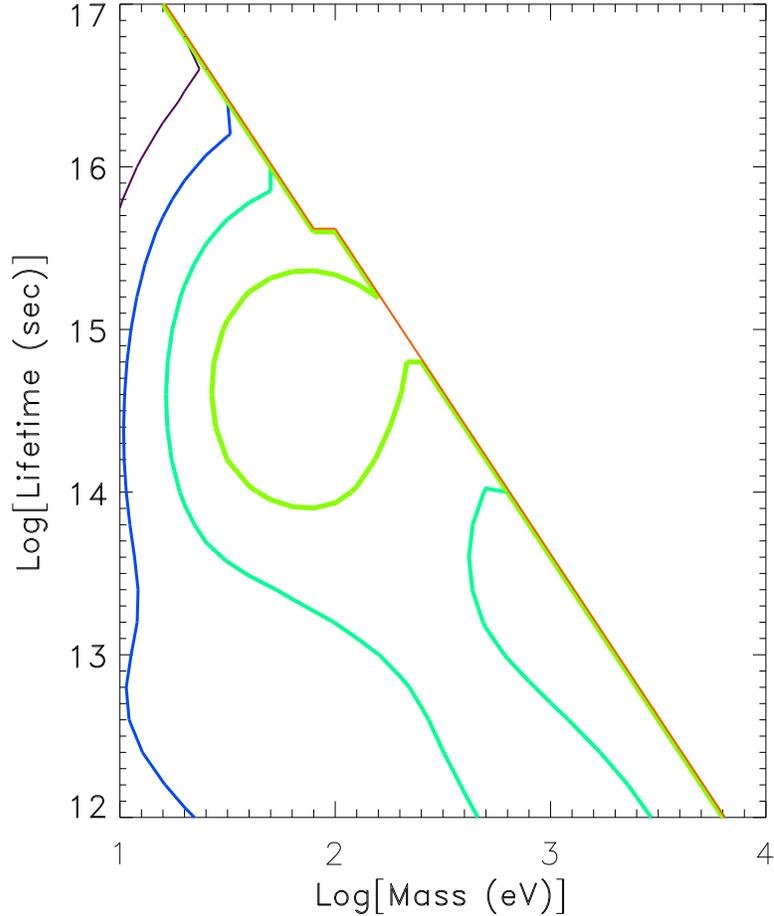}\\
\
\caption[Fig4]{\label{contone}  Same as Fig.~\ref{cont},
but now $\Omega_\Lambda=0.7, \Omega_{total} = 1$.}
\end{figure}

We checked that the contours for a different set of $(h,\Omega_b)$ were
similar to the contours in Fig~\ref{cont}. Fig.~\ref{contone} shows the results
for a cosmological constant-dominated universe. Again, a sizable region is
ruled out, reflecting the robustness of the constraint.

Hannestad~\cite{hannestad} performed a similar calculation, using future CMB
experiments to rule out decaying neutrino models, but he used the same approximation
as in reference~\cite{lopez}; the decay products were added into the background
neutrino density. We expect that his excluded-region contours for low masses should
shrink since ISW effect is the main discriminator for these masses.

% \textbf{Hannestad~\cite{hannestad} performed a more detailed treatment of
% the neutrino decay process than in reference \cite{lopez}, but he used the same
% approximation for the CMB calculations; i.e., the decay products were added into the
% background neutrino density.  However, it is difficult to say what changes, if any,
% would result in his conclusions using our improved treatment, since he compared with
% the much more accurate future CMB data (which means his results rely on much smaller
% variations in the CMB fluctuation spectrum). But we expect that his excluded-region
% contours for low masses should shrink since ISW effect is the main discriminator for
% these masses.}
%
 
% \begin{comm}... below are some comments about structure formation constraints.
% \end{comm}

It has been noted that the decay products from a very massive neutrino could keep the
universe substantially populated with radiation or even radiation-dominated for most
of its history. The presence of radiation has the effect of stopping the growth of
density perturbations, which in a matter-dominated universe would grow as $\delta
\sim a$. Since these density perturbations should (eventually) collapse into the
structure we see today, it is clear that structure formation arguments can also
provide constraints on the neutrino mass and lifetime. Very coarse constraints on the
radiation density can placed by requiring that the scales relevant to structure
formation are able to grow sufficiently (assuming of course, we know the initial
perturbations), as is done in Ref.~\cite{gary+mike}.
In fact, for a scale-invariant
initial spectrum, the structure formation arguments of Ref.~\cite{gary+mike} also
rule out a region at the bottom-right of our excluded region. A more detailed analysis
yields more stringent constraints
\cite{bhar+seth}. In light of this, it is important to understand that the
constraints from CMB are most useful for low masses, i.e., for massive decaying neutrinos
which do not affect the late-time growth of the density perturbations
appreciably. Future experiments (MAP and PLANCK) have the potential to constrain
neutrino masses as low as 1 eV and maybe even lower~\cite{hannestad}. In the end, CMB
and large scale structure constraints on massive decaying neutrinos both overlap and
complement each other.

\section{Conclusions}

Our results indicate that for calculations involving the effects of decaying
particles on CMB fluctuations, exact conservation of energy-momentum (not just
conservation of the mean energy-momentum) is crucial.  When perturbations in the
decay products are correctly treated as being determined by the perturbations in both
the metric and the decaying massive particle, energy and momentum of the massive
particle plus its decay products are conserved. The result is a much smaller change
(when an unstable neutrino is added) in the CMB fluctuation spectrum than was noted
in ref. \cite{lopez}.  However, by using a comparison with current data, rather than
a simple constraint on $C_{200}/C_{10}$, we have been able to obtain an excluded
region only slightly less restrictive than that obtained in ref. \cite{lopez}.  This
excluded region will grow as more data becomes available, culminating
potentially in very restrictive limits from MAP and PLANCK \cite{hannestad}.
Our results, of course, can be generalized to arbitrary decaying particles.

The CMB spectra used in this work were generated with a
modified version of CMBFAST~\cite{selj+zald}. We thank
Lloyd Knox for providing the data used to generate the constraints
in section IV.
This work was supported by the DOE and the NASA grant NAG 5-7092
at Fermilab 
and by the DOE grant DE-FG02-91ER40690 at Ohio State. 

\appendix

\section{Source of Decaying Potentials}

Here we show that the ISW effect in the decaying neutrino model
is primarily driven by the dipole of the decay-produced radiation,
$\theta_{rd}$. The ISW effect is driven by time changes to the potentials
(eq. 2), which are determined from Einstein's equations.
In synchronous gauge, the source of these time changes is
\be
\dot{\phi} + \dot{\psi} = t_1 + t_2 + t_3 + t_4 \,,
\ee
where\\
\parbox{1cm}{\begin{eqnarray*}\end{eqnarray*}}
\hfill
\parbox{2in}{\begin{eqnarray*}
t_1 &=& \left[2+\frac{3}{k^2}\adotoasq(5+3w)\right] \dot{\eta} \,,\\
t_3 &=& - 2 \adotoa \eta \,,
\end{eqnarray*} }
\parbox{2in}{\begin{eqnarray*}
t_2 &=& \frac{1}{2k^2}\adotoasq (5+3w) \dot{h} \,,\\
t_4 &=& \frac{3}{k^2} \left(2 \adotoa D_\sigma - \dot{D}_\sigma\right)\,.
\end{eqnarray*} }
\hfill
\parbox{1cm}{\begin{eqnarray}\label{eq:isw}\end{eqnarray}}\\
The quantity $D_\sigma$ is related to the anisotropic stress of the
fluid: $D_\sigma= -(3/2)\,(\dot a/a)^2 \,(1+w) \,\sigma$,
where $w = P/\rho$ is the equation of state of the universe. We will
consider the behavior of superhorizon-scale perturbations, where
$k \tau \ll 1$. We assume that the neutrinos decay well into the
matter-dominated phase of the universe, and that the decay radiation never dominates
the energy density of the universe, but does come to dominate the standard radiation,
i.e., photons and massless neutrinos. Then the equation of state takes a simple form
near neutrino decay: $w \simeq 1/3\,\Omega_{rd}$. 
In addition, the total fluid perturbation sources
$\theta$ and $\sigma$ are dominated by the decay radiation, so that we can write
$\theta \simeq 4w\,\theta_{rd}$, and $\sigma \simeq 4w\,\sigma_{rd}$. These
assumptions are well motivated for $m_h = 10$ eV, $t_d = 10^{15}$ sec neutrinos which
decay well into the matter dominated era, with $\Omega_{rd} \simeq 0.15$ at decay.

We first examine the behavior of \potd in the approximation where we neglect the
$\dot r_{rd} / r_{rd}$ terms in the Boltzmann equations for the decay radiation
perturbations, i.e., we treat the decay radiation as massless neutrinos as in Lopez,
et al~\cite{lopez}. We then consider the effect of relaxing the approximation and
calculating the decay radiation perturbations correctly. We denote the use of the
Lopez et al. \cite{lopez}
approximation in all quantities by the superscript-$a$.

The potentials do not decay in a completely matter-dominated universe; $\dot{\phi} +
\dot{\psi}$ is sourced by the decay radiation and is therefore first order in
$w$. The term $t^a_4$ is directly related to $\sigma$ and so is of order $w$. The
linearized Einstein equations imply that $\dot \eta \propto \theta \propto w \,
\theta_{rd}$, so that $t^a_1$ is also of order $w$. However, $t^a_2$ and $t^a_3$ are
each zeroth order in $w$, so their sum must cancel to lowest order. Using the
linearized Einstein equations and the continuity equation we find that 
\begin{equation}
t^a_2 + t^a_3 \simeq 8 \frac{w\eta}{\tau} \,,
\end{equation}
%\begin{comm}... replaced 8.4 by 8 since we don't want to say the calculation is 
%good to first decimal place (it isn't). \end{comm}
demonstrating the required cancellation. In our approximation, the decay radiation
perturbations can be calculated from the Boltzmann equation for massless neutrinos,
which admit analytic solutions for the superhorizon modes of $\theta^a_{rd}$ and
$\sigma^a_{rd}$. Using these solutions, it can be shown that 
\be
t^a_1 \simeq -29 \frac{w\eta}{\tau} \,,\;
t^a_4 \simeq 12 \frac{w\eta}{\tau} \,,
\label{t1andt4}
\ee
so that $t^a_1$, $t^a_2+t^a_3$ and $t^a_4$ each contribute roughly comparable amounts
to \potd. In calculating the effect of the approximation on \potd we will therefore
have to separately consider each term. The quantities $D_\sigma$ and $\dot
\eta$ are very much affected by the approximation, since they directly depend on the
decay radiation perturbations, and the error in the decay radiation perturbations is
of order the quantities themselves. This implies that $\delta \dot \eta \sim | \dot
\eta |$ and $\delta D_\sigma \sim | D_\sigma |$, where $\delta x\equiv | x-x^a |$ is
the absolute error in the variable $x$. The zeroth order quantities $\eta$ and $\dot
h$ are much less affected by the approximation.  For super-horizon modes, we do not
expect $\eta$ to evolve much from its initial value, and so the error in it
(determined by the error in $\dot \eta$) is naturally small. Following this line of
reasoning, one can write
\begin{eqnarray}
\delta \eta  &\sim&  \delta \dot \eta  \,\tau \sim k^2\,\tau^2\,w\,\eta, 
\label{deltaeta}\\
\delta \dot h  &\sim& k^2\,\tau\,\delta \eta \sim k^4\,\tau^3\, w\, \eta. 
\label{deltahdot}
\end{eqnarray} 
Using these relations we find that
\be
 \delta t_1 \sim \frac{w\,\eta}{\tau} \,,\;
\delta t_2 \sim (k\tau)^2 \frac{w\,\eta}{\tau} \,,\;
\delta t_3 \sim (k\tau)^2 \frac{w\,\eta}{\tau} \,,\;
\delta t_4 \sim \frac{w\,\eta}{\tau} \,,
\label{deltavars}
\ee
%\begin{comm}... turned the last line in the preceeding paragraph into 
%equations (\ref{deltaeta}, \ref{deltahdot}). added definition for $\delta x$.
%\end{comm} 
which makes it clear that for superhorizon modes, the errors in $t_1$ and
$t_4$ dominate the error in \potd. Numerically it is seen that the error in $t_1$ is
the most important.

We can see why the error in $t_1$ is the most important in a simple way. The dominant
source term (see eq.~\ref{thetarddot}) for $\dot{\theta}^a_{rd}$ is $\deltard^a$. Now
$\dot{\delta}_{rd}$ and $\dot{\delta}^a_{rd}$ differ only by the terms on the
right-hand side of eq.~\ref{deltarddot} (since $\dot h$ is not much affected by the
approximation and $\thetard\ll \dot h$ for super-horizon modes).  But the right-hand
side of eq.~\ref{deltarddot} contains the difference of $\delta_h$ and $\deltard$,
and hence the fractional error in $\deltard$ is expected to be much smaller relative
to that in $\thetard$ or $\shearrd$.  So we can assume that $\deltard$ and
$\deltard^a$ are roughly the same for the purpose of estimating the errors in
$\thetard$ and $\shearrd$ in the approximate scheme. Therefore, for super-horizon
modes (at $\tau\sim\tau_d$), we can write to a good approximation
\begin{equation}
\dot{\theta}_{rd} \simeq \dot{\theta}^a_{rd} 
- \frac{\dot r_{rd}}{r_{rd}}\,\theta_{rd}
\quad \mbox{and} \quad
\dot{\sigma}_{rd} \simeq \dot{\sigma}^a_{rd} 
- \frac{\dot r_{rd}}{r_{rd}}\,\sigma_{rd}
\,. \label{good-approx}
\end{equation}  From this equation, we can gauge that the fractional errors in $\thetard$ and
$\shearrd$ are roughly the same, and close to $-\tau\,\dot r_{rd}/r_{rd}$.  But since
the coefficient for $t_4$ in \potd is much less than that for $t_1$ (see
eq.~\ref{t1andt4}), we expect that the error in $t_1$ dominates which implies that
$\delta (\dot \phi + \dot \psi) \propto \dot \eta \propto \theta_{rd}$.  From
eq.~\ref{good-approx} we have that $\theta^a_{rd} > \theta_{rd}$ for super-horizon
modes at $\tau \sim \tau_d$. Therefore for these modes,
\begin{equation}
\left| {\dot\phi}^a + {\dot\psi}^a \right| 
> \left| \dot\phi + \dot\psi \right|
\,.
\end{equation}
This is the reason for the dramatic rise in power at large scales when we neglect the
decay terms in the decay radiation Boltzmann equations. 
%\begin{comm}... can't look at the coefficients of $t_1$ and $t_4$ and say that the 
%$t_1$ error dominates. need to look at the errors in $t_1$ and $t_4$.
%\end{comm}

%There is a pretty good way to neglect the decay radiation clustering...
Merely adding the decay radiation to the massless neutrino background causes large
errors in the ISW effect. However, there exists a method of calculation that yields
good results without introducing a separate Boltzmann hierarchy for the decay
radiation. Since this method might be helpful for other late time processes
which affect the CMB spectrum, and since it demonstrates that we really
have isolated the source of our disagreement with Lopez et al. \cite{lopez},
we present it here.

\begin{figure}[Fig2]
\centering
\leavevmode\epsfxsize=12cm \epsfbox{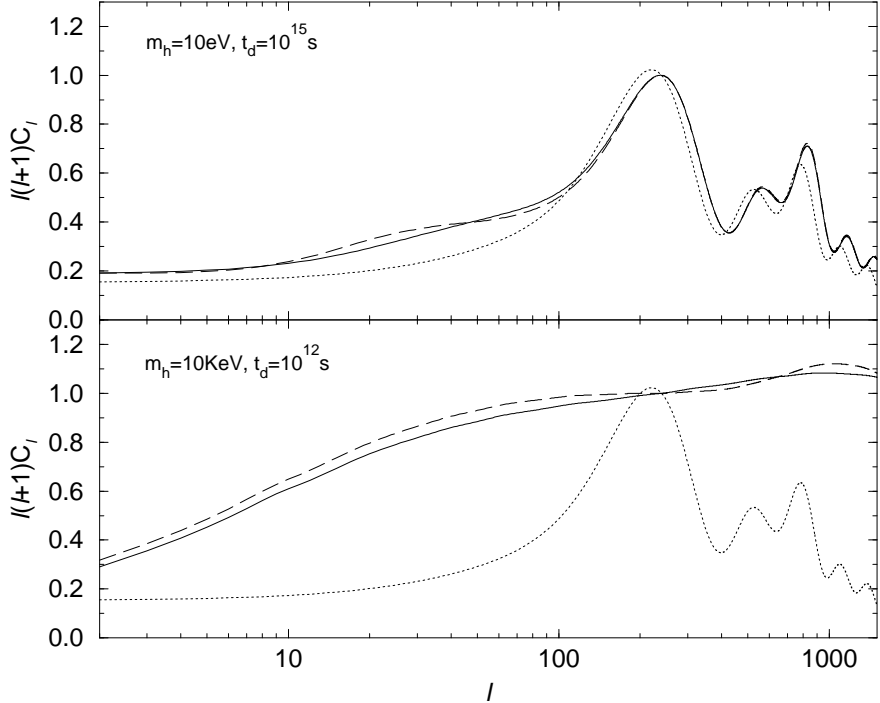}\\
\
\caption[Fig2]{\label{clfix}  The spectrum of CMB fluctuations for
a standard CDM model ($\Omega_b = 0.08$, $h = 0.5$) in the presence of a decaying
neutrino with the indicated mass and lifetime (solid curve).  The dashed curve is the
spectrum obtained when the decay products are added to the background neutrinos and
$\alpha$ is evolved as an explicit differential equation instead of being fixed as in
the standard CMBFAST code.  The dotted curve gives the fluctuation spectrum in the
absence of a decaying neutrino.}
\end{figure}

The fix can be accomplished by explicitly evolving $\alpha = 1/(2k^2)(\dot h
+ 6\dot \eta)$, a quantity that contains the problematic $\dot \eta$, within CMBFAST
in the following differential equation:
\be
%\dot{\alpha}+2\adotoa\alpha=\eta-3D_\sigma.
\dot{\alpha}+2\adotoa\alpha=\eta - \frac{9}{2k^2}\,\adotoasq\,(1+w)\,\sigma
\label{alphadiff}
\ee
The dominant quantity on the right-hand side is $\eta$, which is quite unaffected by
the approximation (recall that $\delta \eta \sim k^2\,\tau^2\,w\,\eta$. In contrast,
when $\alpha$ is set by the equation (the default in CMBFAST),
\be
% \alpha=\frac{a}{\dot{a}}(D_\delta+\eta)+\frac{3D_\theta}{k^2},
\alpha = \frac{a}{\dot{a}}\,\eta + \frac{3}{2 k^2}\,\adotoa \,\delta 
+\frac{9}{2k^4}\,\adotoasq\,(1+w)\,\theta
\,,
\label{alphaset}
\ee
%\begin{comm}... signs of $\delta$, $\theta$ and $\sigma$ terms in 
%eqs.~\ref{alphaset} and~\ref{alphadiff} changed. 
%\end{comm}
the $\theta$ term is important, and inaccuracies in the large scale behavior are
generated. The condition that $\dot{\alpha}$ given by Eq.~(\ref{alphadiff}) should
match that obtained from Eq.~(\ref{alphaset}) is conservation of momentum for the
massive neutrino and its decay products. In the approximate scheme, the unperturbed
quantities for the decay radiation are calculated correctly while the perturbations
in it are set equal to that of the massless neutrino. This violates the
energy-momentum conservation conditions for the system of the massive neutrino plus
its decay products, and this is the reason behind the fact that different
combinations of the Einstein equations lead to different potential decay rates.  It
may be noted that the CMBFAST code~\cite{selj+zald} used to calculate the fluctuation
spectrum implicitly assumes energy-momentum conservation.  When this condition is
violated, the code cannot produce internally consistent results.  In
Fig.~\ref{clfix}, we have plotted the result of using Eq.~\ref{alphadiff} (in place
of Eq.~\ref{alphaset}) with the approximate scheme and as expected, good agreement
with the actual curves is obtained. This exercise clearly shows that it is important
to check for energy-momentum conservation when using approximate methods to model any
part of the energy-momentum tensor.

\begin{thebibliography}{99}
\bibitem{Bunn} E.~F.~Bunn \& M.~White, \apj {\bf 480}, 6 (1997).
\bibitem{deBer} P.~de Bernardis et al., \apj {\bf 480}, 1 (1997).
\bibitem{Lineweaver} C.~H.~Lineweaver, \apj {\bf 505}, L69 (1998).
\bibitem{Hancock} S.~Hancock et al., MNRAS {\bf 294}, L1 (1998).
\bibitem{Les} J.~Lesgourgues et al., astro-ph/9807019 (1998).
\bibitem{Bart} J.~Bartlett et al., astro-ph/9804158 (1998). 
\bibitem{BJ} J.~R.~Bond \& A.~H.~Jaffe, astro-ph/98089043 (1998).
\bibitem{webster} A.~M.~Webster, \apj {\bf 509}, L65 (1998).
\bibitem{white} M.~White, \apj {\bf 506}, 485 (1998).
\bibitem{Ratra} B.~Ratra et al., \apj {\bf 517}, 549 (1999).
\bibitem{MAX} M.~Tegmark, \apj {\bf 514}, L69 (1999).
\bibitem{Liddle}A. R. Liddle, A. Mazumdar \& J. D. Barrow,
\prd {\bf 58}, 027302 (1998);
X. Chen \& M. Kamionkowski, astro-ph/9905368 (1999).
\bibitem{Hannestad}S. Hannestad, astro-ph/9810102 (1998).
\bibitem{alpha}M. Kaplinghat, R.J. Scherrer \& M.S. Turner, 
astro-ph/9810133 (1998).
\bibitem{QED}R. E. Lopez, S. Dodelson, A. Heckler \& M. S. Turner, 
\prl {\bf 82}, 3952 (1999).
\bibitem{stable}
S.\ Dodelson, E.I. Gates, \& A. Stebbins, \apj {\bf 467}, 10 (1996);
J.R.\ Bond, G.\ Efstathiou \& M.\ Tegmark, MNRAS {\bf
291}, L33 (1997).
\bibitem{unstable}
M. White, G. Gelmini \& J. Silk, \prd {\bf 51}, 2669 (1995);
S.~Hannestad, Phys. Lett. {\bf B431}, 363 (1998);
J.A.~Adams, S.~Sarkar \& D.W.~Sciama,  MNRAS {\bf 301}, 210 (1998);
S.~A.~Bonometto \& E.~Pierpaoli, astro-ph/9806035 (1998);
E.~Pierpaoli \& S.~A.~Bonometto,  astro-ph/9806037 (1998).
\bibitem{lopez}R. E. Lopez, S. Dodelson, R. J. Scherrer \& M. S. Turner, 
\prl {\bf 81}, 3075 (1998).
\bibitem{hannestad} S. Hannestad, \prd {\bf 59}, 125020 (1999).
\bibitem{selj+zald}U. Seljak \& M. Zaldarriaga, \apj {\bf 469}, 
437 (1996).
\bibitem{hu+sugy}W. Hu \& N. Sugiyama, \prd {\bf 50}, 627 (1994). 
\bibitem{kawasaki}M. Kawasaki, G. Steigman \& H. Kang, Nucl. Phys. B 
{\bf 403}, 671 (1993).
\bibitem{ma+bert}C.-P. Ma \& E. Bertschinger, \apj {\bf 455}, 7 (1995).    
\bibitem{bond+szal}J. R. Bond \& A. Szalay, \apj {\bf 276}, 443 (1983).
\bibitem{bond+efst}J. R. Bond \& G. Efstathiou, Phys. Lett. B {\bf 265}, 
245 (1991).
\bibitem{gary+mike}G. Steigman \& M. Turner, Nuc. Phys. B {\bf 253}, 
375 (1985).
\bibitem{bhar+seth}S. Bharadwaj \& S. K. Sethi, \apj Supp. {\bf 114}, 
37 (1998).


%%%%%%%%%%
%%%% CMB Experiments
%%%%%%%%%%
\bibitem{cobe} C. Bennett et al., \apj {\bf 464}, L1 (1996).
\bibitem{firs} K. Ganga, L. Page, E. S. Cheng \& S. S. Meyer, \apj
	{\bf 432}, L15 (1994).
\bibitem{tenerife} S. M. Gutteriez de la Cruz et al., \apj {\bf 442}, 10 (1995);
	S. Hancock et al., Nature {\bf 367} 333 (1994); R. Watson et al.,
	Nature {\bf 357} 660 (1992).
\bibitem{bam} G. S. Tucker et al., \apj {\bf 475}, L73 (1997)
\bibitem{southpole} T. Gaier et al., \apj {\bf 398}, L1 (1992); 
	J. Schuster et al., \apj {\bf 412}, L47 (1993);
	J. O. Gundersen et al., \apj {\bf 413}, L1 (1993); 
	J. O. Gundersen et al., \apj {\bf 443}, L57 (1995). 
\bibitem{python} J. Ruhl et al., \apj {\bf 453}, L1 (1995);
		S. R. Platt et al., \apj {\bf 475} L1 (1997).
\bibitem{qmap} M. Devlin et al., astro-ph/9808043 (1998);
	T. Herbig et al., astro-ph/9808044 (1998);
	A. de Oliveira-Costa et al.,astro-ph/9808045 (1998).		
\bibitem{sp89} P. Meinhold \& P. Lubin, \apj {\bf 370},	
	L11 (1989).
\bibitem{argo} S. Masi et al., \apj {\bf 463},
	L47 (1996); P. deBernardis et al., \apj {\bf 422},
	L33 (1994).
\bibitem{max} A. Clapp et al., \apj {\bf 433}, L57 (1994);
	M. J. Devlin et al., \apj {\bf 430}, L1 (1994);
	M. A. Lim et al., \apj {\bf 469}, L69 (1996).
\bibitem{sask} C. B. Netterfield et al., \apj {\bf 474}, 47 (1995);
	E. Wollack et al., \apj {\bf 476}, 440 (1995).
\bibitem{msam} E. Cheng et al., \apj {\bf 422}, L37 (1994);
	E. Cheng et al., \apj {\bf 488}, L59 (1997); 
	G. W. Wilson et al., astro-ph/9902047 (1999).
\bibitem{cat} P. F. S. Scott et al., \apj {\bf 461}, L1 (1996);
	J. C. Baker et al., astro-ph/9904415 (1999).
%%%%%%%%%%
%%%%%%%%%%
\bibitem{BJK} J.R. Bond, A.H. Jaffe \& L.E. Knox, 
astro-ph/9808264 (1998).
\bibitem{TB} S. Burles \& D. Tytler, astro-ph/9803071 (1998).
\bibitem{Press} W. H. Press, S. A. Teukolsky, W. T. Vetterling \&
B. P. Flannery, {\it Numerical Recipes} (Cambridge, 1992).
\bibitem{DK} S.~Dodelson \& L.~Knox, {\it in preparation}.
\end{thebibliography}
\end{document}